\documentclass[twocolumn,prl,showpacs,aps]{revtex4-1}

\usepackage{amsfonts}
\usepackage{amsmath}
\usepackage{amssymb}

\usepackage{graphicx}
\usepackage{dcolumn}
\usepackage{bm}
\usepackage{indentfirst}
\makeatother 
\begin{document}

\title{Magnetic field dependence of tunnel couplings in carbon nanotube quantum dots}
\author{K. Grove-Rasmussen$^{1,2}$}
\email{k\_grove@fys.ku.dk}
\author{S. Grap$^{3}$}
\author{J. Paaske$^{2}$}
\author{K. Flensberg$^{2}$}
\author{S. Andergassen$^{3,4}$}
\author{V. Meden$^{3}$}
\author{H. I. J\o rgensen$^{2}$}
\author{K. Muraki$^{1}$}
\author{T. Fujisawa$^{1,5}$}

\affiliation{$^1$NTT Basic Research Laboratories, NTT Corporation,
3-1, Morinosato-Wakamiya, Atsugi 243-0198, Japan}

\affiliation{$^2$Niels Bohr Institute \& Nano-Science Center,
University of Copenhagen, Universitetsparken 5, 2100~Copenhagen \O,
Denmark}

\affiliation{$^3$Institut f\"{u}r Theorie der Statistischen Physik and
JARA - Fundamentals of Future Information Technology, RWTH Aachen
University, 52056 Aachen, Germany}

\affiliation{$^4$Faculty of Physics, University of Vienna,
Boltzmanngasse 5, 1090 Wien, Austria}

\affiliation{$^5$Department of Physics, Tokyo Institute of
Technology, 2-12-1 Ookayama, Meguro 152-8551, Japan}

\begin{abstract}
By means of sequential and cotunneling spectroscopy, we study the
tunnel couplings between metallic leads and individual levels in a
carbon nanotube quantum dot. The levels are ordered in shells
consisting of two doublets with strong- and weak tunnel couplings,
leading to gate-dependent level renormalization. By comparison to a
one- and two-shell model this is shown to be a consequence of
disorder-induced valley mixing in the nanotube. Moreover, a parallel
magnetic field is shown to reduce this mixing and thus suppress the
effects of tunnel-renormalization.
\end{abstract}

\maketitle
\date{}
Confined states and their tunnel couplings to the leads are the
basis for a wealth of intriguing phenomena observed in quantum dot
nanostructures. In carbon nanotubes the states appear in a
particularly simple arrangement of near four-fold degenerate shells
stemming from spin and orbital degrees of freedom
\cite{Liang:2002,Buitelaar:2002}. The splitting of each quartet into
two doublets is well understood within a single particle model
including disorder-induced valley mixing and spin-orbit interaction
\cite{Kuemmeth:2008,Jespersen:2011}. Less is known about the tunnel
couplings, but experiments show that they can be different for the
two doublets within a shell \cite{Holm:2008}. In a clean nanotube,
such behavior would be inconsistent with time-reversal symmetry.

In this Letter, we show that the relevant doublets indeed may have
different lead couplings due to orbital disorder, and that this
tunnel coupling asymmetry is tunable by a parallel magnetic field
$B_{||}$.
\begin{figure*}[t]
\begin{center}
\includegraphics[width=0.9\textwidth]{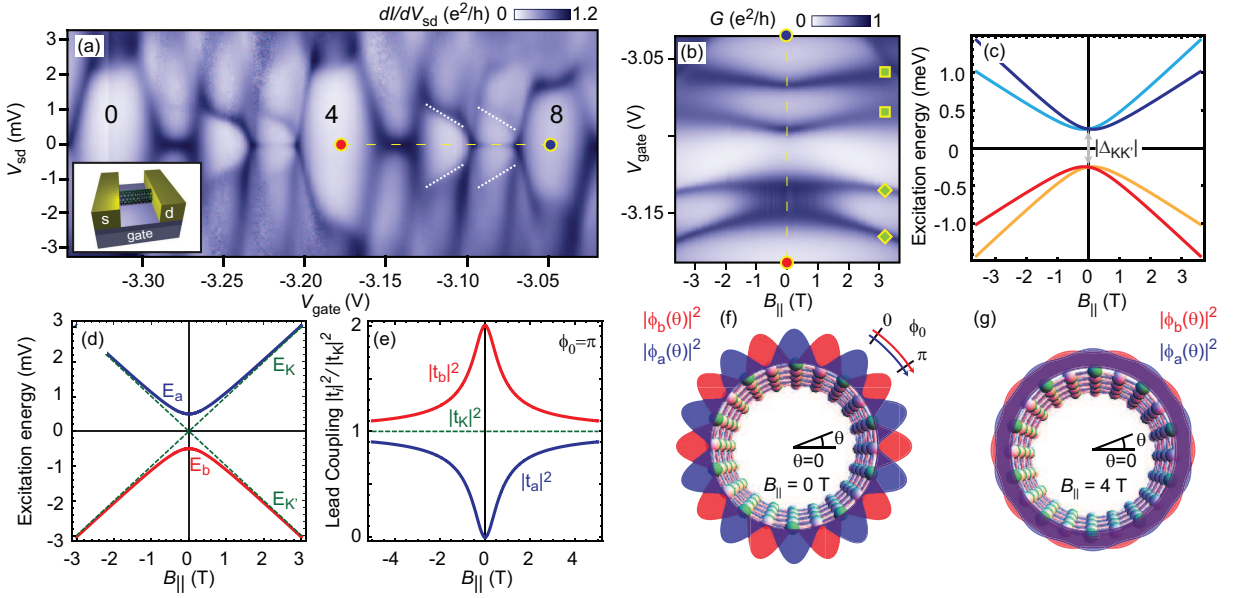}\vspace{-.45cm}
\end{center}
\caption{\label{Oneshell} (color online). (a) Charge stability
diagram of two nanotube shells showing a pattern of strong/weak
Kondo effect for odd occupation and gate-dependent inelastic
cotunneling thresholds. Inset:  backgated two terminal carbon
nanotube device. (b) Linear conductance versus $B_{||}$ for the
$V_{\rm gate}$ region marked by the dashed yellow line in (a). (c)
Shell energy diagram showing the eigenenergies in the presence of an
orbital coupling $\Delta_{KK'}$. (d,e) Level energies and level-lead
couplings of a simple spin-less model. Due to $\Delta_{KK'}$ (with
orbital phase $\phi_0=\pi$), the eigenstates are now differently
coupled to the leads at $B_{||}=0$, while becoming more equally
coupled as $B_{||}$ is increased. (f, g) Simple picture showing that
the electron probability distributions for the two eigenstates are
different/similar at zero/large field. }
\end{figure*}
At $B_{||}=0$\,T, strong and weak Kondo effects in the charge
stability diagrams and gate-dependent inelastic cotunneling
thresholds indicate different doublet couplings \cite{Holm:2008}. At
finite $B_{||}$, Coulomb peaks exhibit $B_{||}$ dependent level
broadenings indicating field tunable tunnel couplings, which is
further corroborated by the disappearance of gate-dependent
inelastic cotunneling lines for equal couplings. A single shell
model including disorder, expressed as an intra-shell coupling
$\Delta_{KK'}$ between the orbital states $K$ and $K'$, is shown to
account well for the observed coupling behavior at low fields. At
larger $B_{||}$, the coupling strengths of the doublets even
interchange and show non-monotonic behavior, which can be explained
by a two-shell model including an inter-shell orbital coupling
$\Delta_{KK'12}$ \cite{Jarillo-Herrero:2005Orbital}. By applying the
functional renormalization group (fRG) to the inner four levels of
the two-shell model, it is shown that the linear conductance versus
gate voltage $V_{\rm gate}$ is in good agreement with the
measurement for all $B_{||}$ \cite{karrasch:2006} including
correlation effects such as Kondo effect.
%

Figure \ref{Oneshell}(a) shows a bias spectroscopy plot obtained by
measuring the differential conductance at temperature $T=140$\,mK of
a 400\,nm long AuPd contacted single wall carbon nanotube device
versus source-drain bias and backgate voltage (see inset in Fig.\
\ref{Oneshell}(a)) \cite{Jespersen:2011}. In the $V_{\rm gate}$
range shown, quantum states belonging to two near four-fold
degenerate shells in the valence band are filled. This leads to the
characteristic pattern \cite{Liang:2002} of three smaller faintly
visible diamonds followed by a big (truncated) diamond as $V_{\rm
gate}$ is increased. Inside each diamond the nanotube holds an
integer number of electrons, $N$, indicated by the additional
electron number for filled shells. For odd occupancy of the shells,
zero-bias conductance ridges are observed, which is a manifestation
of the well-known spin-half Kondo effect
\cite{goldhaber:1998,Nygard:2000}. Interestingly, the Kondo ridges
are broad and narrow for $N+1$ and $N+3$ electrons in the shell,
respectively. A gate-dependent step in conductance at small finite
bias (especially pronounced for filling $N+2$ and $N+3$ in the small
diamonds) reveals the splitting of the four states within a shell
into two doublets via inelastic cotunneling; {\emph i.e.}, at bias
voltages corresponding to the doublet energy difference, an electron
may tunnel into the excited doublet provided that an electron
tunnels out from the ground state doublet \cite{Franceschi:2001}.
Both features can be understood by assuming that the two doublets
have different couplings to the leads: a strongly (weakly) coupled
ground (excited) state results in a strong (weak) Kondo resonance,
while gate-dependent level renormalizations of the doublets are due
to charge fluctuations between the dot and the leads (see SM)
\cite{Holm:2008}. Until now it has remained unclear how the two
doublets could be differently coupled.

To understand the doublet states in more detail, the zero bias
conductance of a single shell is measured as a function of $B_{||}$
and results are shown in Fig.\ \ref{Oneshell}(b) (gate range in
Fig.\ \ref{Oneshell}(a) marked by horizontal dashed line). The
upwards and downwards energy shifts of pairs of Coulomb peaks with
increasing field are consistent with expectations from a simple
nanotube shell model including intra-shell orbital
coupling $\Delta_{KK'}$ 
\footnote{for now neglecting the small spin-orbit interaction} with
energies as shown in Fig.\ \ref{Oneshell}(c)
\cite{Jarillo-Herrero:2005Orbital,Jespersen:2011} (see SM). The
lower (red/orange) and upper (blue/cyan) lines are the resulting
eigenstates corresponding to a finite intra-shell coupling.
Moreover, the widths of the Coulomb peaks (increasing with tunnel
coupling) emerging from the broad Kondo ridge become narrower as the
field is increased (diamonds), while a broadening is observed in
case of the narrow Kondo ridge (squares) \cite{Kondosuppressed}. The
tunnel couplings to the leads are thus observed to be $B_{||}$
dependent.

The observed strong and weak tunnel coupling behavior of the
doublets
is a consequence of the intra-shell orbital coupling as can readily
be understood from a simple spinless model with level energies shown
in Fig.\ \ref{Oneshell}(d) (see SM). Time-reversal symmetry demands
that $|t_K|=|t_{K'}|=t$, but in the presence of a finite intra-shell
orbital coupling, $\Delta_{KK'}=|\Delta_{KK'}|e^{i\phi_0}$, so that
the new eigenstates $|b/a\rangle \sim \mp
e^{i\phi_0}|K\rangle+|K'\rangle$ at $B_{||}=0$ are tunnel coupled by
$t(1\mp e^{i\phi_0})$, respectively (see SM) \cite{Holm:2008}. For
$\phi_0=\pi$ and $B_{||}=0$, one of these states completely
decouples from the leads, while the original eigenstates are
restored and couplings become equal for $B_{||} \gg |\Delta_{KK'}|$
(see Fig.\ \ref{Oneshell}(e)). With this picture in mind, the
decrease (increase) in coupling versus field of the strongly
(weakly) coupled doublets is readily understood. The model gives a
microscopic picture of why the tunnel couplings of the two doublets
can be different. It does not, however, address the issue of
conserved quantum numbers required for the SU(4) or orbital Kondo
effect
\cite{Jarillo-Herrero:2005Orbital,Makarovski:2007Evolution,Makarovski2007}.
Figure \ref{Oneshell}(f) shows that, the electron density (and
therefore tunneling amplitude) of the two doublets may be different
depending on the exact tunneling site $\theta$ on the circumference.
As the field is increased, the respective electron densities become
similar (pattern changes from standing waves to plane waves),
consistent with equal tunnel couplings of the doublets (see Fig.\
\ref{Oneshell}(g) and SM).
\begin{figure}[t!]
\begin{center}
\includegraphics[width=0.5\textwidth]{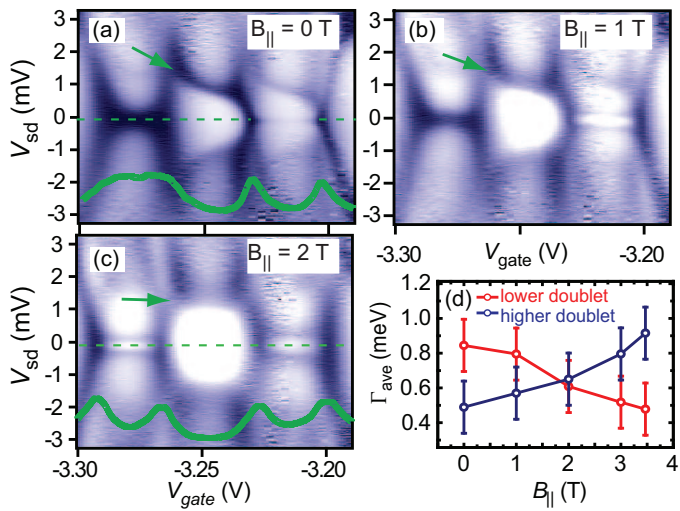}\vspace{-.45cm}
\end{center}
\caption{\label{OneshellHolm} (color online). (a-c) Charge stability
diagrams at $B_{||}=0,1,2$\,T showing that the gate dependence
(originating from differently coupled doublets) of the inelastic
cotunneling threshold vanishes as the parallel field is increased in
accordance with the field dependence of the doublet coupling
asymmetry (see (d)). (d) Extracted lead couplings from Coulomb peak
widths ($\Gamma_\mathrm{ave}$ at $B_{||}=0$ for the lower doublet is
a rough estimate, since the fitting formula is not valid in the
Kondo regime - see SM).}
\end{figure}

\begin{figure*}[t!]
\begin{center}
\includegraphics[width=1\textwidth]{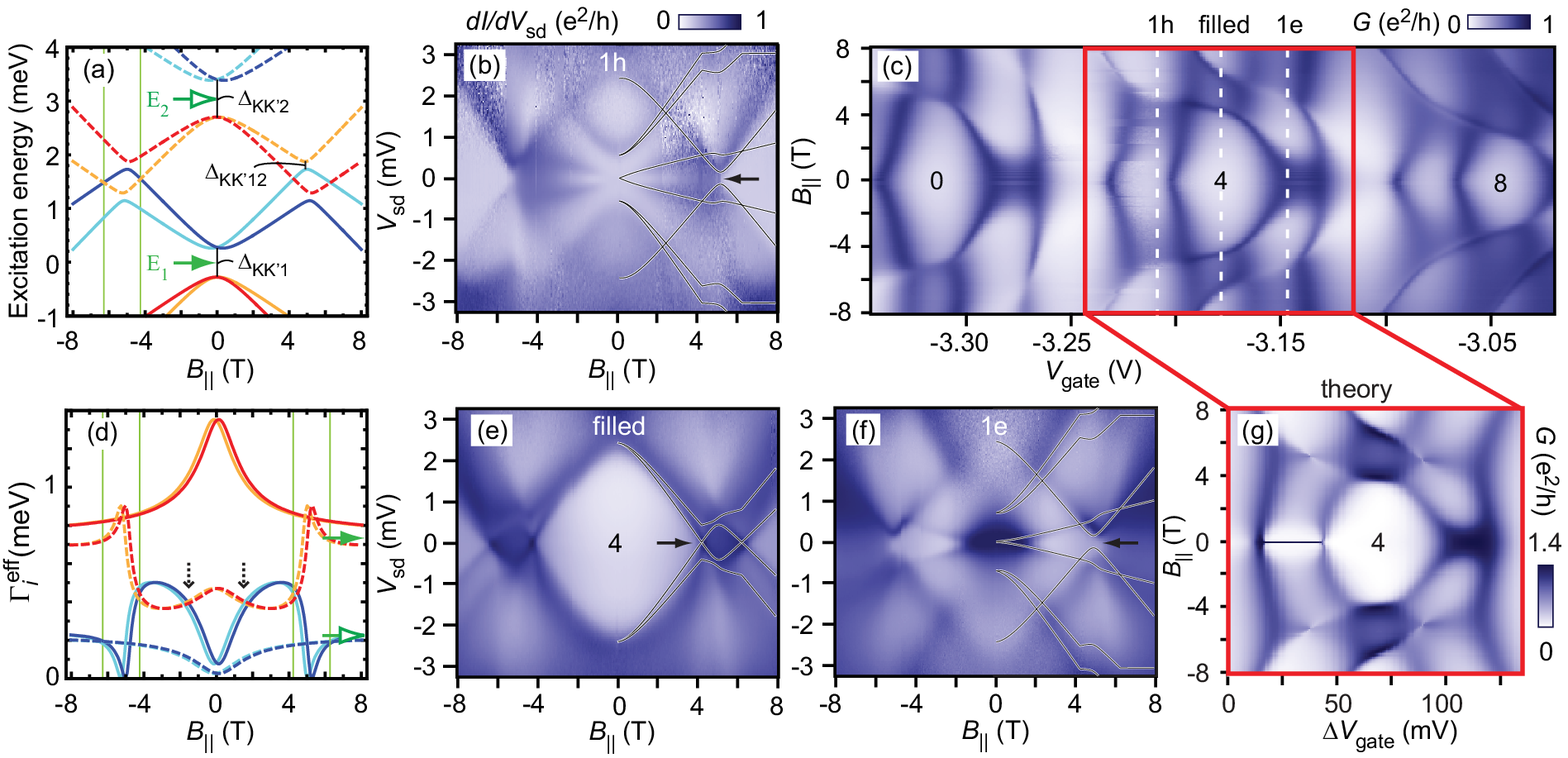}\vspace{-.45cm}
\end{center}
\caption{ \label{Twoshells} (color online). (a) Energy levels of a
two-shell model with level crossings (see vertical lines) resulting
in finite $B_{||}$ Kondo ridges in (c) for levels having different
spins. In contrast, avoided crossings appear at zero/finite $B_{||}$
due to intra/inter-shell orbital couplings for levels having same
spin. (b) Inelastic cotunneling spectroscopy probing level
differences for 1 hole. Lines are fits (to peaks/dips in
$d^2I/dV_{\rm sd}^2$) based on the two-shell model shown in (a). (c)
Linear conductance versus $B_{||}$ and $V_{\rm gate}$ allowing the
study of level crossings of states belonging to different shells.
(d) Orbital couplings give rise to effective tunnel couplings
$\Gamma^{\mathrm{eff}}_i$ of the eigenstates in (a) (horizontal
arrows indicate bare shell coupling strengths). Note that we show
the sum of the couplings to the left and right leads. The small
shift of the couplings within a doublet along $B_{||}$ is due to
spin-orbit interaction. (e-f) As in (b), but for a filled shell and
1 electron. (g) fRG calculation based on the inner four levels of
(a) using the corresponding tunnel couplings of (d)
\cite{Parameters}.}
\end{figure*}


A striking consequence is the tunability of the level
renormalization by $B_{||}$. Figure \ref{OneshellHolm}(a) shows a
zoom in the stability diagram of the left shell in Fig.\
\ref{Oneshell}(a) at $B_{||}=0$ where the onset of inelastic
cotunneling
 processes is clearly observed (see the arrow). The gate dependence of the onset has been explained by
 tunneling renormalization due to a strongly and a weakly coupled doublet \cite{Holm:2008,Grove-Rasmussen:2009}.
 Knowing that
 the intra-shell orbital coupling may induce differently coupled doublets,
 we can now further test both this simple model and the many-body origin of the
 gate dependence. As $B_{||}$ is increased, the
 couplings to the doublets become equal, and as a result the
 inelastic cotunneling threshold is expected to gradually become gate independent.
 This is indeed observed in the stability diagrams
 at finite $B_{||}$, where the gate dependence
 diminishes for $B_{||} = 1$\,T (Fig. 2(b)) and is absent for $B_{||} = 2$\,T (Fig. 2(c)).
The similar Coulomb peak widths in the linear conductance trace at
$B_{||} = 2\,\mathrm{T}$ indicate that the couplings within the
quartet become nearly identical at this field. The $B_{||}$
dependence of the doublet couplings, estimated from the average
width of the two Coulomb peaks belonging to each doublet, is clearly
revealed in Fig. \ref{OneshellHolm}(d) \cite{Jorgensen:2007} (see
SM).

A more careful analysis of the Coulomb peak widths reveals that the
single-shell model does not fully account for all features. At large
fields ($> 3$\,T), the coupling of the weakly coupled doublet
becomes (even) stronger than that of the strongly coupled doublet
(Fig. 1(b)), while the model predicts equal coupling (see Fig.\
\ref{Oneshell}(e)). To fully understand the experimental behavior of
the tunnel couplings at large fields, a two-shell model taking into
account couplings between levels stemming from different shells must
be invoked \cite{4shellsneeded}.
Figure \ref{Twoshells}(c) shows the linear conductance
versus $B_{||}$ and $V_{\rm gate}$ for the gate range shown in Fig.\
\ref{Oneshell}(a). An almost identical behavior of Coulomb peaks
belonging to the two shells is seen, indicating similar tunnel
coupling doublet asymmetries \cite{Kondosuppressed}. Furthermore, at
larger fields, kinks in the Coulomb peaks and slanted conductance
ridges in the Coulomb blockade for filled shells (e.g. filling 4)
indicate level crossing behavior giving rise to the Kondo effect.

For this strongly coupled device, the level structure is best probed
by inelastic cotunneling spectroscopy. Figures
\ref{Twoshells}(b,e,f) show bias spectroscopy plots versus $B_{||}$
for different fillings of the shells along the white dashed lines in
Fig.\ \ref{Twoshells}(c), i.e., one hole (3 electrons), a filled
shell and one electron in the next shell. The observed cotunneling
features are understood by considering two nanotube shells separated
by a level spacing $\Delta E=E_2-E_1$, where in addition both shells
are split into two doublets due to (primarily) intra-shell orbital
interaction $\Delta_{KK'1},\Delta_{KK'2}$ (Fig.\ \ref{Twoshells}(a),
see also SM). As $B_{||}$ is applied, the doublets within each shell
split up, resulting in level crossings between different shells at
elevated fields (see Fig.\ \ref{Twoshells}(a)). The thin black lines
in Figs.\ \ref{Twoshells}(b), \ref{Twoshells}(c), and
\ref{Twoshells}(f) representing cotunneling excitation, which were
obtained from the level spectrum of Fig.\ \ref{Twoshells}(a), well
reproduce the observed features \cite{excitations}. The cotunneling
data also reveal the presence of an inter-shell orbital interaction
$\Delta_{KK'12}$, which couples orbital states with same spin from
different shells, and appears as anticrossings at finite field
(e.g.\ arrows in Figs. \ref{Twoshells}(b) and \ref{Twoshells}(f)).
On the other hand, crossings for states of different spins (e.g.\
arrow in \ref{Twoshells}(f)) indicate that spin-flip scattering is
suppressed \cite{SOIonCouplings}.

The orbital couplings not only modify the internal level structure,
but also induce $B_{||}$ dependent tunnel couplings as shown in
Fig.\ \ref{Twoshells}(d).
These effective $\Gamma$'s are calculated from the eigenstates of
the two-shell model using the bare tunnel amplitudes. At zero field
the doublets arrange in strongly and weakly coupled doublets, but as
the field is increased, the couplings non-monotonically approach the
original couplings of the shells (see horizontal arrows - the
valence band shell at larger negative $V_{\rm gate}$ is more
strongly coupled as often seen in nanotubes). The non-monotonic
coupling behavior is also observed in the data of Fig.\
\ref{Twoshells}(c). For instance the left most Coulomb peak in the
red square initially broadens and then becomes very narrow at high
fields in agreement with Fig.\ \ref{Twoshells}(d) (solid blue line).
It is also seen that the tunnel couplings of the inner two doublets
in Fig.\ \ref{Twoshells}(a) interchange at low fields (see dotted
vertical arrows in Fig.\ \ref{Twoshells}(d)) qualitatively
consistent with Fig.\ \ref{Twoshells}(c), where the widths of the
narrowest Coulomb peaks (two leftmost in the red square) become the
broadest as the field is increased. Whereas the quantitative details
of Fig.\ \ref{Twoshells}(d) depend on the phases of the orbital
couplings (see SM), the agreement with the experiment is better
assessed by comparing the overall features to
an fRG calculation based on the level structure and tunnel couplings.

The fRG calculated linear conductance (see SM) \cite{karrasch:2006}
including inter/intra-shell orbital couplings, spin-orbit couplings,
and the charging energy $U_c$ is shown in Fig.\ \ref{Twoshells}(g)
(compare to square in Fig.\ \ref{Twoshells}(c)). Only the relevant
and correctly modeled inner four levels are kept. After choosing
appropriate intra-shell coupling phases the calculation reproduces
the weak and strong Kondo resonances, the non-monotonic
magnetic-field dependent Coulomb peak widths, as well as Kondo
ridges in Coulomb blockade at finite fields. Interestingly, the
finite field Kondo ridges are seen to be gate dependent, thus giving
rise to a $V_{\rm gate}$ controlled spin ground state transition
\cite{Hauptman:2008,Roch:2008} consistent with different tunnel
couplings of the states involved (see SM) \cite{Grap:2011}. Note
that in the $T=0$ calculation left-right asymmetric level-lead
couplings (corresponding to a higher linear conductance than in the
experiment) were chosen to anticipate a suppression at finite $T$
\cite{frgnote}.

In conclusion, we have shown that the doublets formed in a nanotube
shell in presence of disorder-induced valley mixing may have
different tunnel couplings to the leads. Furthermore, this
difference is modified by applying a parallel magnetic field. The
linear conductance fRG results for a two-shell model show good
agreement with experiments.

We are grateful to P. E. Lindelof and acknowledge help from N.
Kumada, Y. Tokura, J. R. Hauptmann and T. S. Jespersen.
This work was supported by FOR 912 of the DFG (SG, SA, and VM).

\bibliography{CNTrenorm}

\begin{thebibliography}{10}
\bibitem{Kondosuppressed}
The narrow Kondo ridges were suppressed in $B_{||}$ sweeps due to a slightly
  increased $T$ and insufficient resolution.
\bibitem{Parameters}
Shell parameters are fixed by the fitting \{$\Delta E, |\Delta_{KK'1}|,
  |\Delta_{KK'2}|, |\Delta_{KK'12}|, \Delta_{\rm SO}\} = \{3.05, 0.55,$ $0.7,
  0.15, 0.075$\}\,meV and $g_{\rm orb}=5.1$. Right/left bare couplings
  \{$\Gamma_1^L$, $\Gamma_1^R$, $\Gamma_2^L$, $\Gamma_2^R$\}=\{0.56, 0.17,
  0.17, 0.05\}$\,\mathrm{meV}$, orbital coupling phases \{$\phi_1$, $\phi_2$,
  $\phi_{12}$\}=\{$0.85\pi$, $0.85\pi$, $0$\} and $U_c=4\,\mathrm{meV}$ are
  used in the fRG calculation.
\bibitem{4shellsneeded}
A four-shell model is needed to account for the crossing behavior (see SM), but
  the effect on tunnel couplings in a multi-shell systems is best illustrated
  in the simpler case of a two-shell model.
\bibitem{excitations}
The excitations are calculated from the energy difference between the ground
  and excited states in Fig.\ \ref{Twoshells}(a).
\bibitem{frgnote}
The narrow $B_{||}=0$ Kondo ridge has a small Kondo temperature and is
  suppressed in the experiment.
\end{thebibliography}
\bibliographystyle{PRLsep2008}
\end{document}